\pdfoutput=1

\documentclass[aps,prd,nofootinbib,amsmath,amssymb,superscriptaddress,10pt]{revtex4}

\usepackage{amssymb,amsmath,graphicx,color,microtype}

\usepackage{setspace}

\def \der{{\rm d}}
\def \Mp{M_{\rm pl}}

\begin{document}

\title{Reconstructing the Local Potential of Inflation with BICEP2 data}
\author{Yin-Zhe Ma}
\email{mayinzhe@phas.ubc.ca} \affiliation{Department of Physics
and Astronomy, University of British Columbia, 6224 Agricutural Road, 
Vancouver, BC, V6T 1Z1, Canada.} \affiliation{Shanghai Astronomical Observatory,
Chinese Academy of Science, 80 Nandan Road, Shanghai, China,
20003}
\author{Yi Wang}
\email{yw366@cam.ac.uk} \affiliation{Centre for Theoretical
Cosmology, DAMTP, University of Cambridge, Wilberforce Road, Cambridge, CB3 0WA, UK}

\begin{abstract}
We locally reconstruct the inflationary potential by using the
current constraints on $r$ and $n_{\rm s}$ from BICEP2 data.
Assuming small and negligible $\alpha_{\rm s}$, the inflationary
potential is approximately linear in $\Delta\phi\sim \Mp$ range
but becomes non-linear in $\Delta\phi\sim 10 \Mp$ range. However
if we vary the value of $\alpha_{\rm s}$ within the range given by
constraints from {\it Planck} measurement, the local
reconstruction is only valid in the range of $\Delta\phi\sim 0.4
\Mp$, which challenges the inflationary background from the point
of view of effective field theory. We show that, within the range
of $\Delta \phi \sim 0.4 \Mp$, the inflation potential can be
precisely reconstructed. With the current reconstruction, we show
that $V(\phi) \sim \phi^{2}$ and $\phi^{3}$ are consistent, while
$\phi$ model is ruled out by $95\%$ confidence level of the
reconstructed range of potential. This sets up a strong limit of
large-field inflation models.


\end{abstract}

\maketitle

\section{Introduction}
\label{sec:intro} The Inflation paradigm \cite{Guth81,Linde82} is
successful in explaining the horizon problem, flatness problem and
the homogeneity problem in the standard hot-big-bang cosmology.
The generic inflation model predicts a nearly scale-invariant
primordial scalar power spectrum which has been measured
accurately by the observations of the cosmic microwave background
radiation (CMB) such as {\it Wilkinson Microwave Anisotropy Probe}
(hereafter {\it WMAP}) \cite{Hinshaw13} and {\it Planck}
\cite{Planck16} satellites. However, even with precise constraints
from CMB temperature fluctuations, there are still many models
that predict the values of spectral index $n_{\rm s}$ and its
running $\der n_{\rm s}/\der \ln k$ which are allowed by the
constraints from current data.

Recently, the ground-based ``Background Imaging of Cosmic
Extragalactic Polarization'' experiment just completed its second
phase experiment (hereafter BICEP2), which observed the CMB B-mode
polarization (divergence-free mode of polarization) on angular
scales of a few degrees \cite{BICEP2} (For cosmological
implications, see also \cite{cosmoa,cosmob,cosmoc,cosmod}). The
CMB B-mode polarization can only be sourced by primordial
gravitational waves, which is a very clean test of the primordial
tensor fluctuations. Results from BICEP2~\cite{BICEP2} show that
the power spectrum of B-mode polarization $C^{BB}_{\ell}$ on a few
degree angular scales is detected at $\sim 7\sigma$ confidence
level (CL), which clearly indicates a signature of primordial
gravitational waves. If this is true, it becomes a strong
observational support of the scenario in which the Universe
started from the inflationary exponential expansion, when the
primordial tensor fluctuations are produced and stretched to
super-Hubble length, and later entered into the Hubble horizon and
decayed at small scales.

Indeed, this field of CMB observation has been developing very
fast over the past decades and many on-going experiments are
seeking such a CMB B-mode polarization signal. For instance, the
{\it Planck} satellite with its nine frequency channels may
achieve higher signal-to-noise ratio and probe even larger angular
scales than BICEP2. Ground-based SPTPol \cite{Austermann12},
ACTPol \cite{Niemack10}, PolarBear \cite{PolarBear} and CLASS
\cite{Eimer12} experiments are also completing with each other to
make more precise measurement on the CMB B-mode polarization
signals. Therefore further experiments may precisely determine not
only the amplitude but also the shape of the primordial tensor
power spectrum, therefore constitutes a direct test of the
inflation mechanism.

Therefore it is important to connect the predictions from
inflation models with the current observational results from
BICEP2 and {\it Planck}. In pervious {\it WMAP} and {\it Planck}
analysis papers \cite{Hinshaw13,Planck22}, the authors plot the
predictions of spectral index of scalar power spectrum $n_{\rm s}$
and tensor-to-scalar ratio $r$ of various inflation models with
the constraints from CMB data (fig.~7 in \cite{Hinshaw13} and
fig.~1 in \cite{Planck22}). While making the prediction of $n_{\rm
s}$--$r$ relation for a given potential, the variation of the
inflaton field is calculated by integrating the equation of motion
from the end of inflation to some early epoch. This duration of
inflation is assumed by to around $50$--$60$ number of e-folds
($N=\log(a/a_{\rm i})$). Although the $n_{\rm s}$-$r$ relation
works well, it is worth noticing the underlying assumption that
during inflation, the inflaton potential (which is typically taken
as a monomial, for example, $V\propto \phi^2$) is the same as that
during the first 10 e-folds of observable inflation.

With the recent measurement of tensor-to-scalar ratio $r$, this
assumption become problematic. It becomes much more challenging
than before to build an inflation model, in which a simple
potential describes the total $60$ e-folds of inflation without
changing its shape and parameters. To see this, remember that the
inflationary potential can be perturbatively expanded near a value
of $\phi_*$ as
\begin{eqnarray}\label{eq:eft}
  V(\phi) =  V(\phi_*) + \partial_\phi V \Delta\phi +
  \cdots   \frac{1}{4!} \partial_\phi^4 V \Delta\phi^4 + \cdots ~,
\end{eqnarray}
where $\Delta \phi =\phi- \phi_{\ast}$ is the change of $\phi$
value during inflation. From the effective field theory point of
view, the potential derivatives up to $\partial_\phi^4 V$ are
relevant and marginal operators. Those operators can be naturally
turned on without suppression. On the other hand, the
$\partial_\phi^4 V$ and higher derivatives are irrelevant
operators, which are suppressed with an energy scale defined by
the UV physics (at most the Planck scale). For the expansion
\eqref{eq:eft} to converge we need $\Delta\phi$ to be smaller than
the UV completion scale of inflation. However, Lyth bound
\cite{Lyth:1996im} suggests that, the change of the field with
respect to the number of e-folds is related to the value of $r$
\begin{eqnarray}
  \left|\frac{\der \phi}{\der N}\right| = \frac{\Mp}{4}\sqrt{2r},
  \label{eq:Lyth}
\end{eqnarray}
where $\Mp=(8 \pi G)^{-1/2}$ is the reduced Planck mass. By
substituting the current measurement of $r$ from BICEP2
\cite{BICEP2}
\begin{equation}
  r = 0.20 ^{+0.07}_{-0.05} \text{  } (1\sigma{\rm CL}). \label{eq:r-val}
\end{equation}
Thus per e-fold, $\Delta\phi = 0.16 \Mp$. By assuming $N \simeq
60$, we find that the inflaton field moves at least at a distance
\footnote{Here we do not take the time variation of $\epsilon$
into account, to avoid model dependence. Otherwise the number in
\eqref{eq:phi60} could change, while keep within the same order of
magnitude. Also note that it is also possible that $\epsilon$ is
not varying monotonically, to avoid large field inflation
\cite{Hotchkiss:2011gz,Ben-Dayan10}.}
\begin{align} \label{eq:phi60}
  \Delta\phi \simeq  9.6 \Mp ~,
\end{align}
in its field space. If this is true, $\Delta\phi$ at 60 e-folds is
much greater than $\Mp$. Thus the expansion \eqref{eq:eft} is
no-longer valid since all the high derivatives of $V$ could in
principle contribute along the 60 e-folds of the inflationary
trajectory. The effective field theory of inflationary background
is therefore non-perturbative, and becomes out of control for
higher order derivatives.

The UV completion of inflation becomes a sharper problem then
ever before. However, the leading UV completion paradigm, string
theory, actually makes the problem worse. On the one hand, most
string inflation models predict much smaller $r$ and thus not
consistent with the BICEP2 data. On the other hand, the
characteristic energy scale of string theory is the string scale.
For string theory to be perturbatively solvable, strong coupling
had better to be small and the string scale should be lower than
the Planck scale (say, 0.1 $M_\mathrm{pl}$ or lower). The size of
extra dimension may further lower the string scale. With such a
lower scale as the cutoff, the effective field becomes a greater
challenge than that with the Planck scale cutoff.

Before BICEP2, the major challenge for building stringy inflation
models is the $\eta$-problem \cite{Copeland:1994vg}, with the
observational $\eta$ smaller than theoretical expectations. Now, a
more serious $\epsilon$-problem emerges, leaving the observed
large $\epsilon$ for the string theorists to explain.

In the effective field theory point of view, given the current
constraint on $r$, we may not be able to trust the inflaton
potential along the whole 60 number of e-folds. This motivates us
not to integrate the potential throughout $60$ number of e-folds,
but to reconstruct the potential \cite{Copeland:1993jj,
Lidsey:1995np,Ben-Dayan10} locally. Therefore we focus on a local
range of field values, along the first a few
 e-folds window. In this range, $\Delta \phi \sim \Mp$ thus the
inflationary potential expanded by Eq.~(\ref{eq:eft}) is in better
control. We will show that, assuming small running,
with current data it is possible to
accurately reconstruct the amplitude and shape of the inflaton
potential within the CMB observation window of about 10 e-folds.
However, in the case
of large running, the uncertainty of the reconstruction becomes large when
$\Delta\phi$ is comparable with $0.4\Mp$, which corresponds to a field range of about 3 e-folds.

This paper is organized as follows: in Sec.~\ref{sec:slow-roll},
we explain our notations of slow-roll parameters, and show the
connection with $n_{\rm s}$ and $r$. In Sec.~\ref{sec:recon-slow}
we directly constrain the slow-roll parameters with current data
from BICEP2. In Sec.~\ref{sec:recon-poten}, we sample the
inflationary potential and compare its amplitude and shape with
the large-field inflation models. The conclusion and discussions
are presented in the last section.

\section{Slow-roll parameters}
\label{sec:slow-roll} The slow-roll parameters as derivatives of
the scale factor can be defined as
\begin{align}
  \epsilon = - \frac{\dot H}{H^2}~,\quad
  \eta = \eta_1 = \frac{\dot\epsilon}{H\epsilon}~,\quad
  \eta_n = \frac{\dot\eta_{n-1}}{H \eta}\quad (n\geq 2)~ \label{eq:def}.
\end{align}
Note that these are not equivalent to the parameters defined by the
derivatives of the inflationary potential. However, the
definitions (\ref{eq:def}) are increasingly commonly used and
their role to keep track of the expansion history of inflation is
by itself important.

At the leading order of slow roll, the scalar power spectrum can be written as  \cite{Chen:2010xka, Wang:2013zva}
\begin{align}
  P_\zeta = \frac{H^2}{8\pi^2 \epsilon \Mp^2}  ~,
\end{align}
where $\Mp =1/\sqrt{8\pi G} = 2.435\times 10^{18}$GeV is the
reduced Planck mass. The spectral index of primordial power
spectrum, the running of the spectral index and the
tensor-to-scalar ratio for single-field slow-roll inflation is
\begin{align} \label{eq:nsm1}
  n_{\rm s} -1 = -2\epsilon -\eta ~,
\end{align}
\begin{align}
  \alpha_{\rm s} = -2\eta \epsilon - \eta \eta_2 ~,
\end{align}
\begin{align} \label{eq:r16e}
  r = 16 \epsilon ~.
\end{align}

Given the current measurement of $r$ (Eq.~(\ref{eq:r-val})) with
$P_\zeta = 2.4 \times 10^{-9}$ at $k_{0}=0.05 {\rm Mpc}^{-1}$
(best-fitting value constrained by {\it Planck}~\cite{Planck22}),
the best-fitting values of $\epsilon$ and Hubble parameter $H$ are
\begin{align}
  \epsilon = 0.0125 ~, \quad H = 4.4 \times  10^{-5} \Mp = 1.1 \times 10^{14}
  \mathrm{GeV},~
\end{align}
respectively. This Hubble scale sets the energy scale of
inflationary perturbations. The corresponding energy density is
\begin{align}
\rho = 3\Mp^2H^2 = 2.0\times 10^{-9}\Mp^4 = (1.6\times
10^{16}\mathrm{GeV})^4~,
\end{align}
which is about $10^{12}$ times higher than the current Large
Hadron Collider (LHC) experiment. Therefore, the CMB experiment is
essentially a high-energy experiment that probes the regime of
physics unaccessible by the current ground-based accelerators.

Note that $\rho^{1/4}$ is of order the maximal temperature
that the universe could get at reheating. The current preferred value $\sim 10^{16}\mathrm{GeV}$ is
interestingly near the grand unification scale. It thus becomes
increasingly important to understand the relation between
inflation and the grand unification, including model building,
reheating mechanism, and possible topological defects which might
be produced at the grand unification phase transition.

The tensor power spectrum and its spectral index are
\begin{eqnarray}
  P_\mathrm{T} = \frac{2H^2}{\pi^2 \Mp^{2}} = 4.8 \times 10^{-10}~, \\
  n_{\rm T} = -2 \epsilon = - \frac{r}{8}  = - 0.025~, \label{eq:nT}
\end{eqnarray}
where the second equation holds only for single-field slow-roll
inflation models.

\section{Reconstruction of the slow-roll parameters}
\label{sec:recon-slow}

\begin{figure*}
\centerline{
\includegraphics[width=0.42\textwidth]{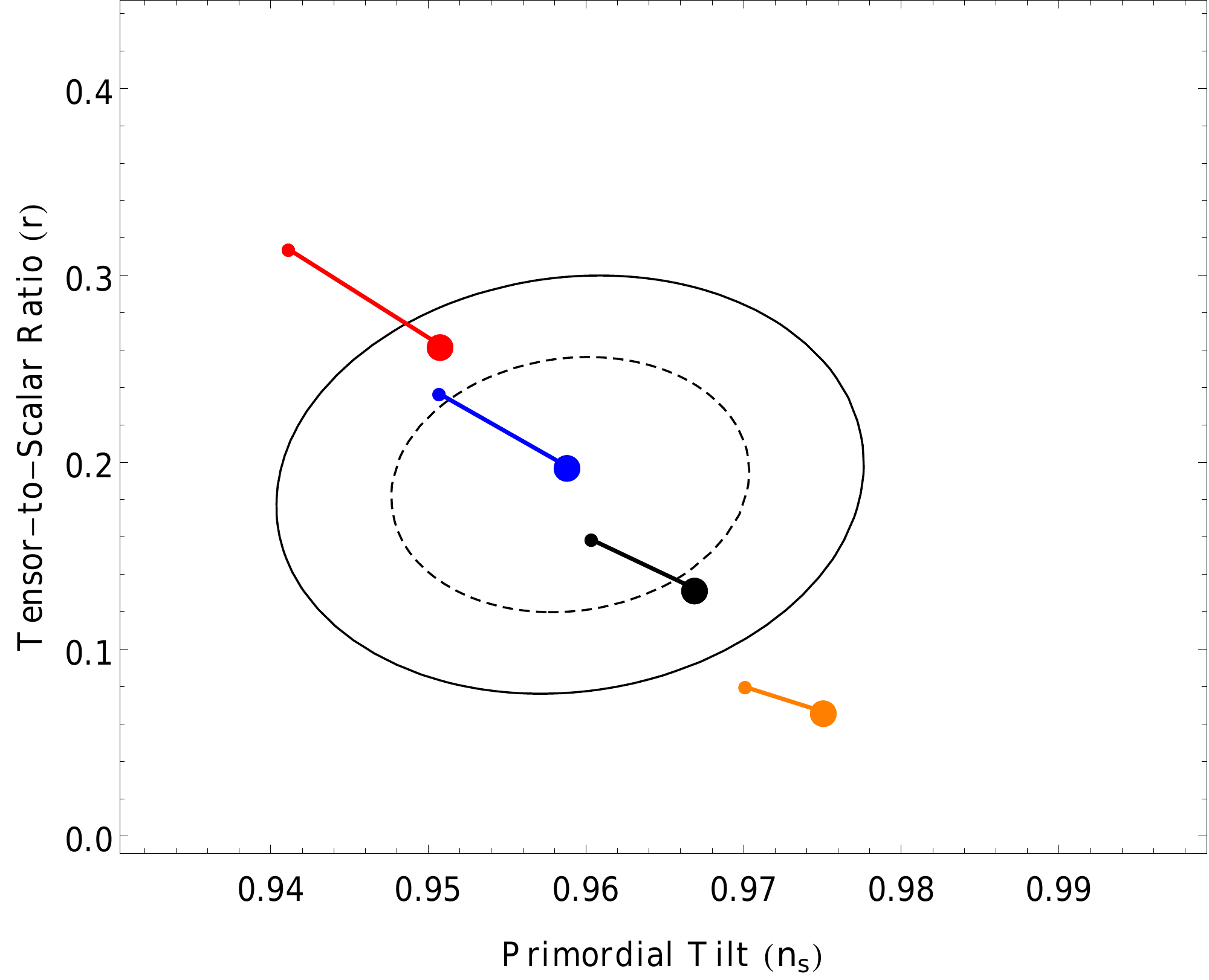}\hspace{0.5cm}
\includegraphics[width=0.435\textwidth]{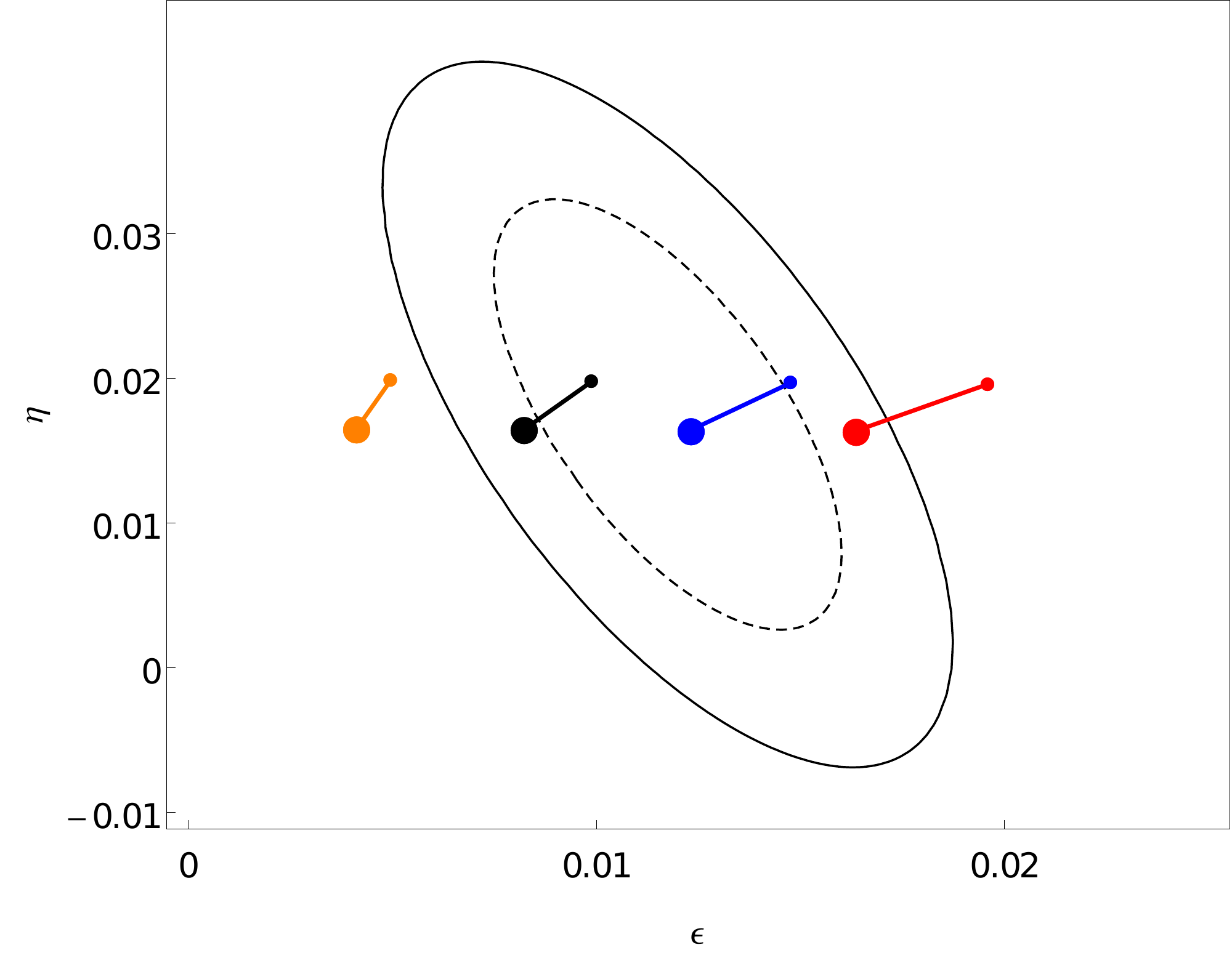}\hspace{0.3cm}
\includegraphics[bb=0 -50 50 300, width=0.03\textwidth]{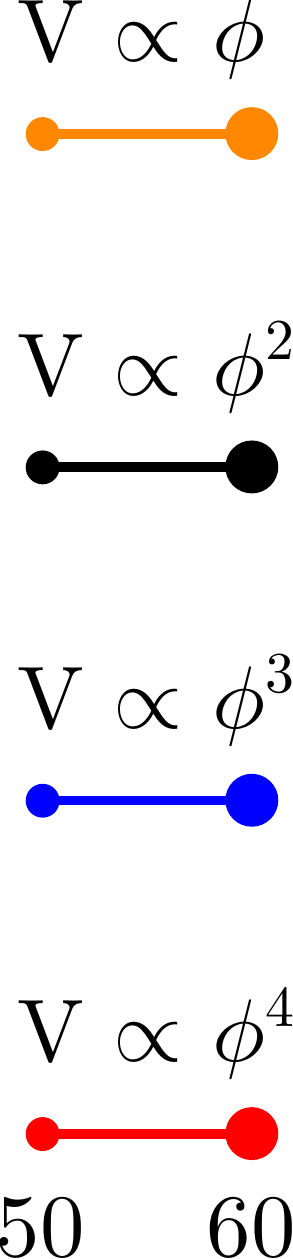}}
\caption{{\it Left}-- Joint constraint on $n_{\rm s}$ and $r$ from
{\it Planck}+WP+highL+BICEP2 data with running \cite{BICEP2}. {\it Right}-- The
derived equal-probability contours on the $\epsilon$-$\eta$ plane.
In both two panels, the dashed and solid lines are the $68.3\%$
and $95.4\%$ confidence level respectively. Note that the $V\propto \phi^n$ models do not have running. (For comparison of those models with the data contours without assuming running, see Fig. \ref{fig:pbcombine}.)}\label{fig:nsr}
\end{figure*}

Figure~\ref{fig:nsr} shows the joint constraints ($1,2\sigma$
confidence level) on $n_{\rm s}$--$r$ with current {\it Planck}+
WP \footnote{This is the {\it WMAP} polarization data
\cite{Hinshaw13}.}+ highL \footnote{This high-$\ell$ CMB data is
mainly from 150GHz South Pole Telescope (SPT)~\cite{Keisler11} and
148GHz Antacama Cosmology Telescope (ACT)~\cite{Das11}.} + BICEP2.
We have also plotted together the prediction of $n_{\rm s}$--$r$
relation during the number of e-folds $N=50$--$60$, for the
large-field inflation models $V \propto \phi^n$, (n=1,2,3,4) (see
Sec.~\ref{sec:compare} for details). We can see that the $\phi^2$
and $\phi^3$ are within or near the $68.3\%$ CL (depending on
e-folds), and the previously considered ``ruled out'' $\phi^4$
potential is now back inside $95.4\%$ contour if $N\simeq 60$. The
linear potential becomes disfavored by the new data.

The $n_{\rm s}$-$r$ diagram can be fitted by the multivariate
normal distribution
\begin{eqnarray}
 L(n_{\rm s},r)= \frac{1}{2\pi \sqrt{1-\rho_{nr}^2} \sigma_n \sigma_r} \exp \left\{
  - \frac{1}{2(1-\rho_{nr}^2)} \left[
    \frac{(n_{\rm s}-1-\mu_n)^2}{\sigma_n^2} + \frac{(r-\mu_r)^2}{\sigma_r^2}
    - 2\frac{(n_{\rm s}-1-\mu_n)(r-\mu_r)\rho_{nr}}{\sigma_n \sigma_r} \right]
  \right\} ~,
\end{eqnarray}
where $\mu_n$, $\sigma_{n}$ ($\mu_r$, $\sigma_{r}$)
are the central value and standard deviation of $n_{\rm s}-1$
($r$) respectively. The $\rho_{nr}$ is their correlation
coefficient. Fitting this multi-variant Gaussian distribution with
the ($n_{\rm s}$, $r$) diagram (Fig.~\ref{fig:nsr}), we find
\begin{align} \label{eq:nrval}
  \mu_n = -0.041 ~, \quad \sigma_n = 0.0075 ~, \quad
  \mu_r = 0.19 ~, \quad \sigma_r = 0.045 ~, \quad \rho_{nr} = 0.10 ~.
\end{align}

The equal-probability contours of multivariate distribution are those with the exponent
\begin{align}
    e_{nr} \equiv \frac{1}{(1-\rho_{nr}^2)} \left[
    \frac{(n_{\rm s}-1-\mu_n)^2}{\sigma_n^2} + \frac{(r-\mu_r)^2}{\sigma_r^2}
    - 2\frac{(n_{\rm s}-1-\mu_n)(r-\mu_r)\rho_{nr}}{\sigma_n \sigma_r} \right] = \mathrm{constant}~.
\end{align}
The $e_{nr}$ as a random variable obeys $(\chi_2)^2$ distribution (i.e. the $\chi^2$ distribution with two degrees of freedom). A contour with probability $\alpha$ inside the contour corresponds to
\begin{align}
  e_{nr}
  = 2\log\left( \frac{1}{1-\alpha} \right)~.
\end{align}


From \eqref{eq:nsm1} and \eqref{eq:r16e}, the inflationary
slow-roll parameters $\epsilon$ and $\eta$ satisfies the
multivariate normal distribution
\begin{align}
L(\epsilon,\eta)= \frac{1}{2\pi \sqrt{1-\rho_{\epsilon\eta}^2}
\sigma_\epsilon \sigma_\eta} \exp \left\{
  - \frac{1}{2(1-\rho_{\epsilon\eta}^2)} \left[
    \frac{(\epsilon-\mu_\epsilon)^2}{\sigma_\epsilon^2} + \frac{(\eta-\mu_\eta)^2}{\sigma_\eta^2}
    - 2\frac{(\epsilon-\mu_\epsilon)(\eta-\mu_\eta)\rho_{\epsilon\eta}}{\sigma_\epsilon \sigma_\eta} \right]
  \right\}~,
\end{align}
with the central value, standard deviation and the correlation
coefficient of $\epsilon$ and $r$ being
\begin{align}
  \mu_\epsilon = \frac{\mu_r}{16}~, \quad
  \sigma_\epsilon = \frac{\sigma_r}{16}~, \quad
\end{align}
\begin{align}
  \mu_\eta = -\mu_n - \frac{\mu_r}{8} ~, \quad
  \sigma_\eta = \frac{1}{8}\sqrt{64\sigma_n^2+16\rho \sigma_n \sigma_r
  +\sigma_r^2}~,
\end{align}
\begin{align}
  \rho_{\epsilon\eta} = \frac{\left(\rho _{nr}^2-1\right) \sigma _r-\rho _{nr} \left| 8 \sigma _n+\rho
   _{nr} \sigma _r\right| }{\sqrt{16 \sigma _n \rho _{nr} \sigma _r+64 \sigma
   _n^2+\sigma _r^2}}.
\end{align}
Plugging in the data from \eqref{eq:nrval}, we obtain
\begin{align}
  \mu_\epsilon = 0.012 ~, \quad \sigma_\epsilon = 0.0028 ~, \quad
  \mu_\eta = 0.018 ~, \quad \sigma_\eta = 0.0098 ~, \quad \rho_{\epsilon\eta} = -0.65 ~.
\end{align}

The $68.3\%$ and $95.4\%$ CL of the joint constraints $\epsilon$
and $\eta$ are plotted in the right panel of Fig.~\ref{fig:nsr}.
One can see that the constraints on large-field inflation is the
same as the left panel of Fig.~\ref{fig:nsr}: the $\phi^{2}$ and
$\phi^{3}$ models are favoured by the current data within
$2\sigma$ CL, while $\phi^{1}$ model is ruled out at
$2$--$3\sigma$ CL.


The probability distribution of $\eta_2$, on the other hand, can
be derived from the current bound of $\alpha_{\rm s}$. Current
data shows \cite{Planck22,Abazajian14}
\begin{align}
  \alpha_{\rm s} = -0.022 \pm 0.010 \quad (68\% \mathrm{CL})~ \label{eq:alphas}.
\end{align}
To good precision, one can approximate $\eta_2 = - \alpha_{\rm s}
/ \eta$, considering that $2\eta\epsilon$ is much smaller than the
current experimental bound.

On the other hand, for most slow-roll models of inflation,
$\eta_2$ is much smaller than the current bound. Here we shall
assume $\eta_3$ and $\eta_4$ are of the order $\eta$  or smaller,
while consider both cases of large $\eta_2$ and small $\eta_2$,
motivated by observations and theory respectively.

\section{Reconstruction of the Inflationary Potential}
\label{sec:recon-poten}
\subsection{Derivatives of the Inflationary Potential}
\label{sec:derivative}

\begin{figure*}[htbp]
  \centerline{
  \includegraphics[width=0.45\textwidth]{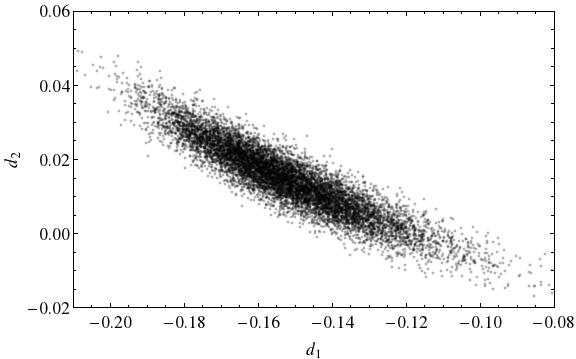}
  \includegraphics[width=0.45\textwidth]{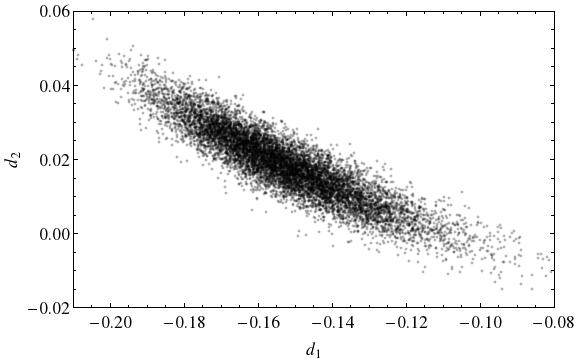}}
  \caption{The probability
distribution of $d_1$ and $d_2$. Here $10^4$ points are dropped to
illustrate the distribution in both panels. {\it Left}-- Assuming
that $\alpha_{\rm s}$ is small and negligible. {\it Right}--
Treating $\alpha_{\rm s}$ as a free parameter and using the
observational constraint $\alpha_{\rm s}$ (Eq.~(\ref{eq:alphas}))
to reconstruct $d_{1}$--$d_{2}$. As one can find (and analytically
expect), the dependence on $\alpha_{\rm s}$ is weak for $d_1$ and
$d_2$.}\label{fig:d1d2}
\end{figure*}


In this section we expand the inflationary potential in terms of
the slow-roll parameters. Since we are only interested in the
range of a few number of e-folds, we are able to locally expand
the potential in an effective field theory and have more confident
to drop the non-renormalizable terms. In addition, higher order
derivatives on the potential are highly suppressed by slow-roll
parameters (and by the largeness of $\Mp$) practically (unless the
higher order slow-roll parameters are unusually huge). Thus we
derive the derivatives of the potential up to $4$th order.

In single field inflation (without slow-roll approximation), the
derivatives of the potential can be solved from the slow-roll
parameters as
\begin{eqnarray}
  \partial_\phi V &=&  \frac{1}{2} H ( - 6 + 2 \epsilon  - \eta )
  \dot\phi, \\ \label{eq:v1}
  \partial^2_\phi V & =& -\frac{1}{4} H^2 \left(8 \epsilon ^2 - 2 \epsilon  (12 + 5 \eta ) + \eta  \left(6 + \eta  + 2 \eta
  _2\right)\right), \\ \label{eq:v2}
  \Mp^2\partial^3_\phi V &=&  \frac{H \dot\phi  \left(8 \epsilon ^3 - 6 \epsilon ^2 (4 + 3 \eta )
 + \epsilon  \eta  \left(18 + 6 \eta  + 7 \eta _2\right)
 - \eta  \eta _2 \left(3 + \eta  + \eta _2 + \eta _3\right)\right)}{4 \epsilon
 }, \\ \label{eq:v3}
 \Mp^2 \partial^4_\phi V &=&   -4 H^2 \epsilon ^3 + 2 H^2 \epsilon ^2 (6 + 7 \eta )
 - \frac{1}{4} H^2 \epsilon  \eta  \left(72 + 39 \eta  + 32 \eta _2\right)
\nonumber\\
 &&  +  \frac{1}{8} H^2 \eta  \left(6 \eta ^2 + \eta  \left(18 + 35 \eta _2\right)
 + 6 \eta _2 \left(8 + 3 \eta _2 + 3 \eta _3\right)\right)
\nonumber\\
 &&  +  \frac{H^2 \eta  \eta _2 \left(\eta ^2 - \eta  \left( - 3 + 3 \eta _2 + \eta _3\right)
 - 2 \left(\eta _2^2 + 3 \eta _2 \left(1 + \eta _3\right)
 + \eta _3 \left(3 + \eta _3 + \eta _4\right)\right)\right)}{8 \epsilon
 } \label{eq:v4}
\end{eqnarray}
Slow-roll approximation simplifies the above equations. However,
it is important to note that if we allow large running, $\eta_2$
could be as large as $\mathcal{O}(1)$. Thus here we perform
slow-roll approximation, but leaves $\eta_2$ not approximated
\footnote{The calculation of
$P_\zeta$, $n_{\rm s}-1$ and $\alpha_{\rm s}$ around local
potential does not rely on the smallness of $\eta_2$. See, for
example, \cite{Wang:2013zva}, for the computational details. This is an advantage of making use of the slow roll parameters defined from expansion. On the other hand, if the slow roll parameters are defined by derivatives of the potential, the large $\eta_2$ enters the calculation, though eventually cancelled in calculating the observables.}. By
using $\dot\phi=\Mp H \sqrt{2\epsilon}$,
Eqs.~(\ref{eq:v1}--\ref{eq:v4}) can be simplified as
\begin{align} \label{eq:dn}
  d_0 &\equiv \frac{V}{3\Mp^2H^2} \simeq 1~, \qquad
\\
  d_1 &\equiv \frac{\Mp}{V}\partial_\phi V \simeq  -\sqrt{2\epsilon} ~,
\\
  d_2 &\equiv \frac{\Mp^2}{V}\partial^2_\phi V \simeq  \frac{1}{2}  (4 \epsilon  - \eta - \frac{1}{3} \eta \eta_2) ~,
\\
  d_3 &\equiv \frac{\Mp^3}{V}\partial^3_\phi V \simeq  - \frac{ \left(8 \epsilon ^2 - 6 \epsilon \eta (1+7\eta_2/18)  + \eta  \eta _2 (1+ (\eta+\eta_2+\eta_3)/3 )\right)}{\sqrt{2\epsilon} } ~,
\\
  d_4 &\equiv  \frac{\Mp^4}{V}\partial^4_\phi V \simeq  \frac{\eta  \eta _2 \left(-2 \eta _2^2+\eta  (\eta +3)-64
   \epsilon ^2+(35 \eta +48) \epsilon +3 \eta _2 \left(-\eta -2
   \eta _3+6 \epsilon -2\right)-\eta _3 \left(\eta +2 \eta _3+2
   \eta _4-18 \epsilon +6\right)\right)}{24 \epsilon }
  \nonumber\\ &
  ~~~~~~~~~~~~~~~~~~~ + \frac{\epsilon  \left(3 \eta ^2+16 \epsilon ^2-24 \eta  \epsilon
   \right)}{4\epsilon}~,
\end{align}
where dimensionless parameters $d_i$ ($i=0,1,2,3,4$) are defined
to measure the derivatives of the inflationary potential. Without
loss of generality we have used $\dot\phi = \sqrt{2\epsilon}\Mp H
>0$, i.e. we have chosen the positive sign solution
instead of the negative sign solution $\dot\phi =
-\sqrt{2\epsilon}\Mp H <0$. This is because given a potential with
a $\dot\phi<0$ solution, one can always flip the potential by
$\phi \rightarrow -\phi$ redefinition without change of any
physics.

With the definition in \eqref{eq:dn}, $d_n$ is of order
$\mathcal{O}(\epsilon, \eta_i)^{n/2}$ in slow-roll parameters.
However, one should note that there may be a hierarchy between
$\epsilon$ and $\eta$ such that the above counting
($\mathcal{O}(\epsilon, \eta_i)^{n/2}$) may break down if
$\epsilon \ll \eta$. Fortunately, with the current tensor we
should have at least $\epsilon \sim \eta$. Thus the slow-roll
order counting should be fine unless fine tuning happens.

With the above definition, the potential can be reconstructed till
$4$th order as
\begin{align} \label{eq:Vrec}
  V(\phi) \simeq V(\phi_*) \left[ 1 + d_1 \left( \frac{\Delta\phi}{\Mp}\right) + \frac{1}{2} d_2 \left(\frac{\Delta\phi}{\Mp}\right)^2 +  \frac{1}{6}d_3 \left(\frac{\Delta\phi}{\Mp}\right)^3 +  \frac{1}{24} d_4 \left(\frac{\Delta\phi}{\Mp}\right)^4 \right]~.
\end{align}
Here one can see explicitly that if $\Delta\phi > \Mp$, the
coefficients $d_n$ are required to be smaller for higher orders in
order for the Taylor expansion to be converged.

\subsection{Sampling the Inflationary Potential}
\label{sec:sample-poten}
 Before performing the numerical reconstruction of potential, we derive a few analytical
relations. Given the distribution of $\epsilon$ and $\eta$, the
statistical properties of $d_1$ and $d_2$ can be calculated as
\begin{align}
  \mu_1 & \equiv \langle d_1\rangle = -\frac{i
   \left(\sqrt{2} \Gamma \left(\frac{3}{4}\right) \sigma _{\epsilon } \,
   _1F_1\left(-\frac{1}{4};\frac{1}{2};-\frac{\mu _{\epsilon }^2}{2 \sigma _{\epsilon }^2}\right)-2
   \Gamma \left(\frac{5}{4}\right) \mu _{\epsilon } \, _1F_1\left(\frac{1}{4};\frac{3}{2};
   -\frac{\mu
   _{\epsilon }^2}{2 \sigma _{\epsilon }^2}\right)\right)}{2^{3/4} \sqrt{\pi } \sqrt{\sigma
   _{\epsilon }}}
\nonumber\\ &
~~~~~~~~~~~~ -\frac{\sqrt{\pi } \mu _{\epsilon }^{3/2} e^{-\frac{\mu _{\epsilon }^2}{4 \sigma
   _{\epsilon }^2}} I_{-\frac{1}{4}}\left(\frac{\mu _{\epsilon }^2}{4 \sigma _{\epsilon
   }^2}\right)}{4 \sigma _{\epsilon }}-\frac{\sqrt{\pi } \mu _{\epsilon }^{3/2} e^{-\frac{\mu
   _{\epsilon }^2}{4 \sigma _{\epsilon }^2}} I_{\frac{3}{4}}\left(\frac{\mu _{\epsilon }^2}{4 \sigma
   _{\epsilon }^2}\right)}{4 \sigma _{\epsilon }}-\frac{\sqrt{\pi } \mu _{\epsilon }^{3/2}
   e^{-\frac{\mu _{\epsilon }^2}{4 \sigma _{\epsilon }^2}} I_{\frac{5}{4}}\left(\frac{\mu _{\epsilon
   }^2}{4 \sigma _{\epsilon }^2}\right)}{4 \sigma _{\epsilon }}
\nonumber\\ &
~~~~~~~~~~~~ -\frac{\sqrt{\pi } e^{-\frac{\mu
   _{\epsilon }^2}{4 \sigma _{\epsilon }^2}} \left(\mu _{\epsilon }^2+2 \sigma _{\epsilon }^2\right)
   I_{\frac{1}{4}}\left(\frac{\mu _{\epsilon }^2}{4 \sigma _{\epsilon }^2}\right)}{4 \sqrt{\mu
   _{\epsilon }} \sigma _{\epsilon }}
+\frac{4 \mu _{\epsilon }^{3/2} \, _2F_2\left(\frac{1}{2},1;\frac{5}{4},\frac{7}{4};-\frac{\mu
   _{\epsilon }^2}{2 \sigma _{\epsilon }^2}\right)}{3 \sqrt{\pi } \sigma _{\epsilon
   }}~, \\
  \sigma_1^{2} &\equiv \langle d_1^2\rangle - \mu_1^2 = 2\mu_\epsilon -
  \mu_1^2~, \\
  \mu_2 &\equiv \langle d_2\rangle = 2 \mu_\epsilon -
  \frac{\mu_\eta}{2}~, \\
  \sigma_2^{2} &\equiv \langle d_2^2\rangle - \mu_2^2 = 4 \sigma_\epsilon - 2 \rho_{\epsilon\eta} \sigma_\epsilon \sigma_\eta + \frac{\sigma_\eta^2}{4}~,
\end{align}
where ${}_pF_q$ is the hypergeometric function and $I_\nu$ is the
modified Bessel function of the first kind.

For the calculation of $\mu_2$ and $\sigma_2$, we have assumed
that $\eta_2$ is small and negligible. However, in the following
numerical sampling, we shall consider both possibilities: either
sampling $\eta_2$ from observational bound of $\alpha_{\rm s}$, or
assuming $\eta_2$ is small and negligible.

\begin{figure*}[htbp]
  \centerline{
  \includegraphics[width=0.35\textwidth]{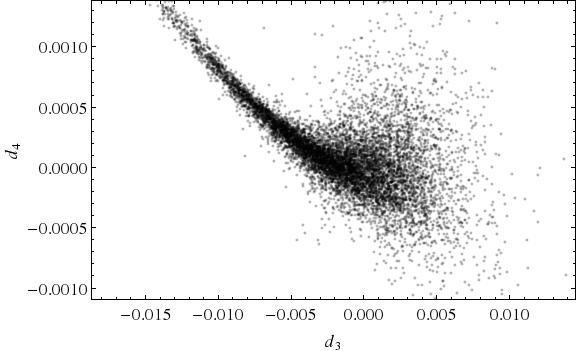}
  \includegraphics[width=0.35\textwidth]{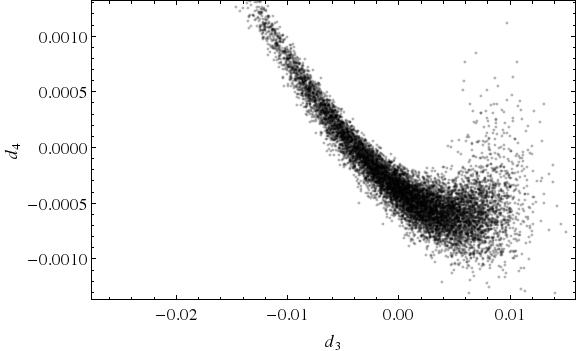}
  \includegraphics[width=0.32\textwidth]{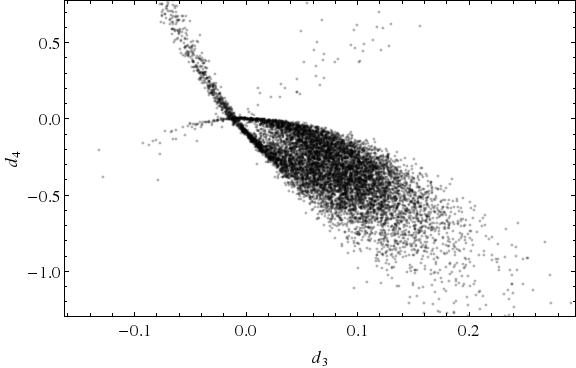}}
  \caption{\label{fig:d3d4} The probability distribution
of $d_3$ and $d_4$. In each panel there are $10^4$ random numbers
chosen to sample the distribution. {\it Left}--$\eta_2$, $\eta_3$
and $\eta_4$ are assumed to be random variables with the same mean
and variance of $\eta$. Unlike the ($d_1$, $d_2$) plot, the
central value of $d_3$ and $d_4$ are around zero while the
distribution is highly distorted. {\it Middle}-- $\eta_2$,
$\eta_3$ and $\eta_4$ set to zero. Comparing with left panel, the
statistical properties are not significantly modified without or
with small $\eta_2$, $\eta_3$ and $\eta_4$. {\it Right}-- $\eta_2$
is determined by the constraint on $\alpha_{\rm s}$
(Eq.~(\ref{eq:alphas})).
  Since the error of $\alpha_{\rm s}$ is still quite significant,
  the parameter
space is broadened by orders of magnitude.}
\end{figure*}

First(and also theoretically reasonable), we can assume that
$\eta_n$ ($n\geq 2$) are random variables with the same variance
as $\eta$ (a difference at the same order of magnitude does not
change the result significantly). The probability distribution of
$d_3$ and $d_4$ are plotted in the left panel of
Fig.~\ref{fig:d3d4}. From the plot, we confirm that with the mild
theoretical assumptions discussed before, the derivative expansion
of the potential converges nicely as a local expansion.
We have also checked our assumption of the $\eta_3$ and $\eta_4$ distribution
in the middle panel of Fig.~\ref{fig:d3d4}, where $\eta_3$ and $\eta_4$ are set
to zero, which does not significantly modify the distribution of $d_3$ and $d_4$.

Second, we also assume non-zero $\alpha_{\rm s}$ and take its
constraints (Eq.~(\ref{eq:alphas})) (the values near the
observational bound) to sample the potential. This possibility
seems not very probable theoretically. But on the other hand,
observationally a large $\alpha_{\rm s}$ would be the easiest way
to resolve the tension between the low $r$ reported by {\it
WMAP}/\textit{Planck}, and the high $r$ reported by BICEP2. The
tension may either be resolved by considering isocurvature
perturbations or the anomalous suppression of power at low $\ell$.
But those possibilities are beyond the scope of the current work, but interested
readers can refer to~\cite{Xia14,Contaldi14}.

The distribution of $d_1$ and $d_2$ is illustrated in
Fig.~\ref{fig:d1d2}, with small $\alpha_{\rm s}$ and observational
$\alpha_{\rm s}$ respectively. As one can find (and analytically
expect), the dependence on $\alpha_{\rm s}$ is weak for $d_1$ and
$d_2$.

In right panel of Fig.~\ref{fig:d3d4}, we use $\alpha_{\rm s}$ to
constrain $\eta_2$. There $\eta_3$ and $\eta_4$ are treated as
having a variance the same as $\eta$. But the choice of $\eta_3$
and $\eta_4$ only affects $d_4$, which is the least important one
in the reconstruction.

\begin{figure*}[htbp]
  \centerline{
  \includegraphics[width=0.31\textwidth]{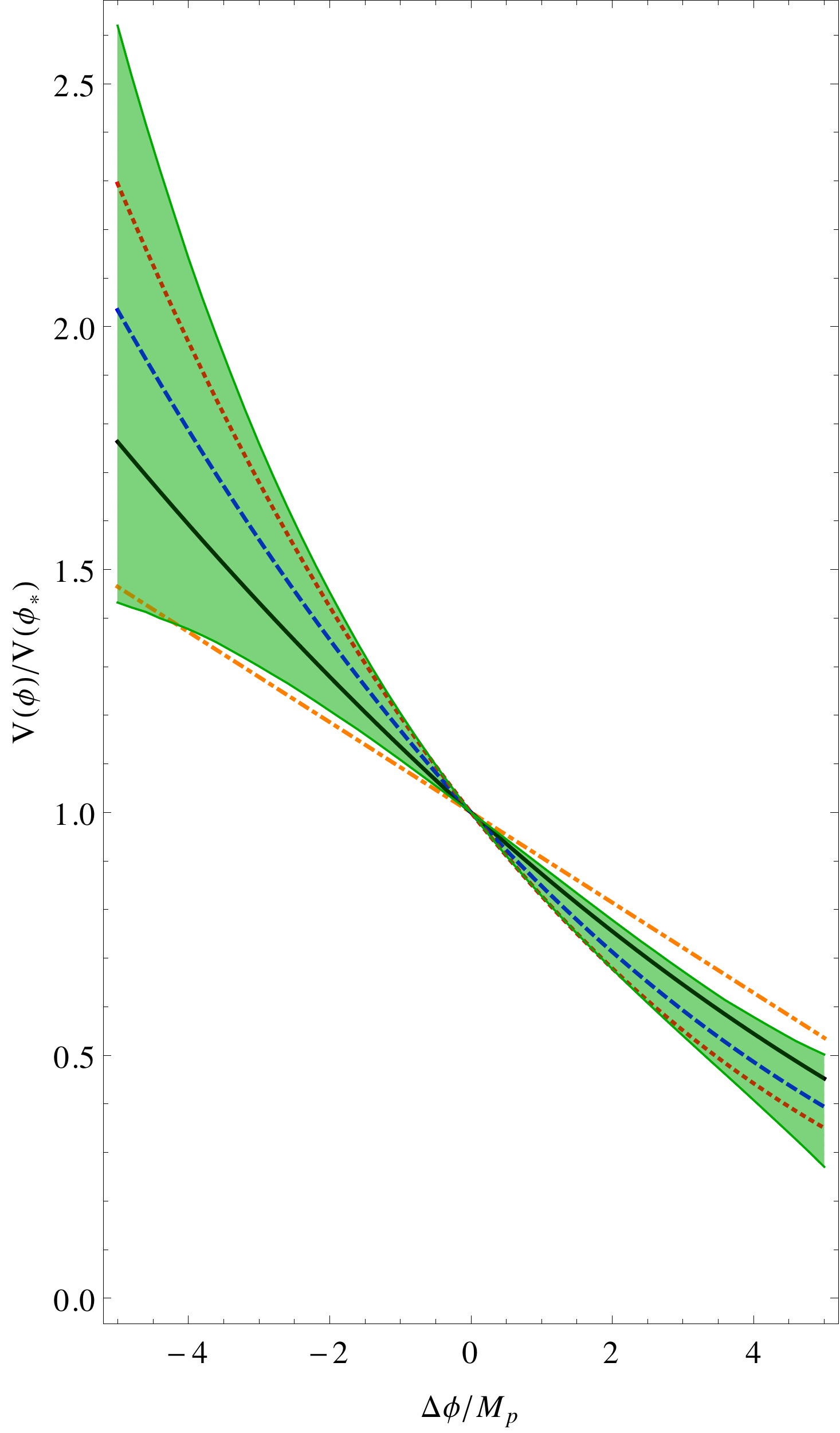}\hspace{0.5cm}
  \includegraphics[width=0.322\textwidth]{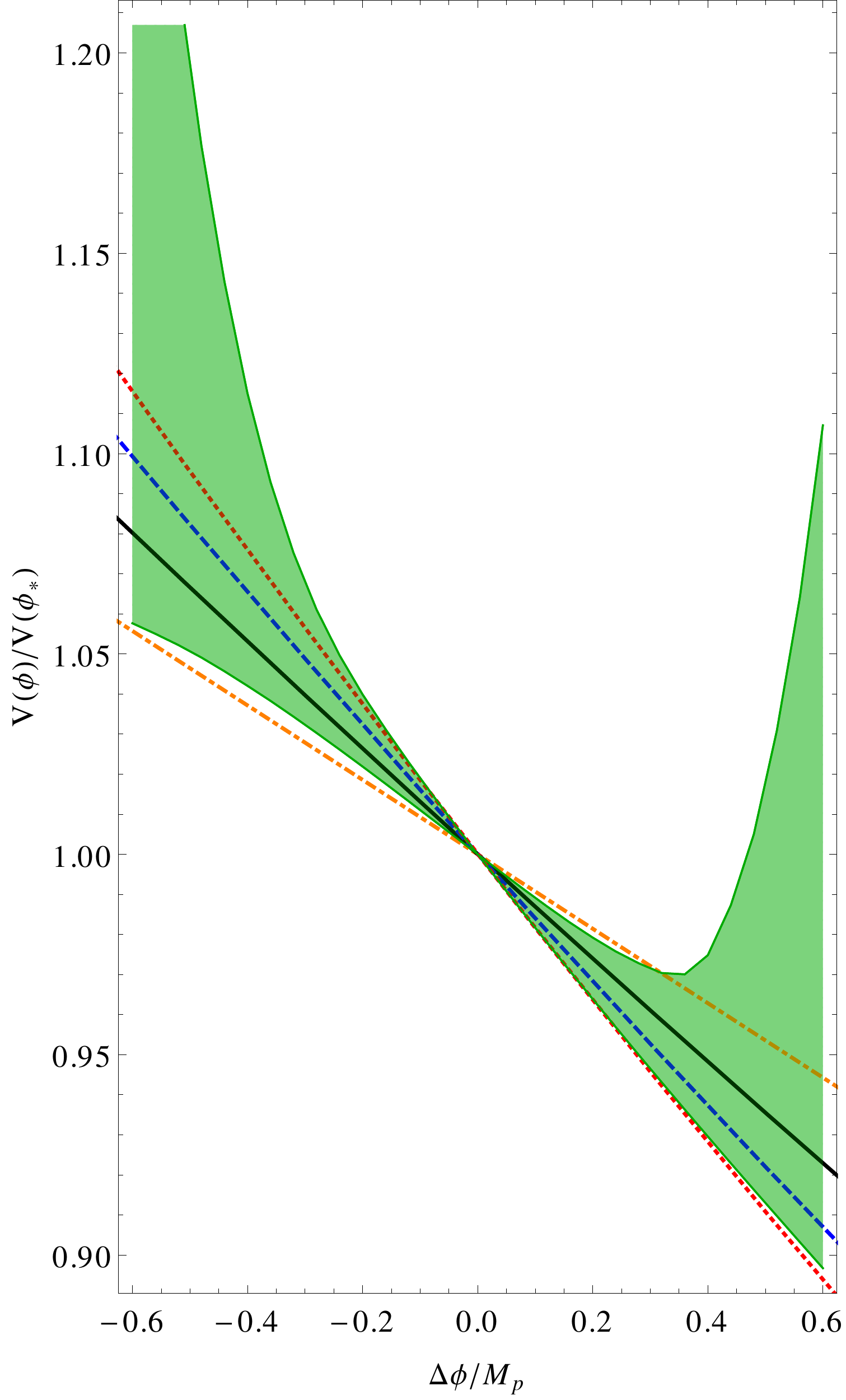}
  \includegraphics[bb=-30 -60 50 300, width=0.07\textwidth]{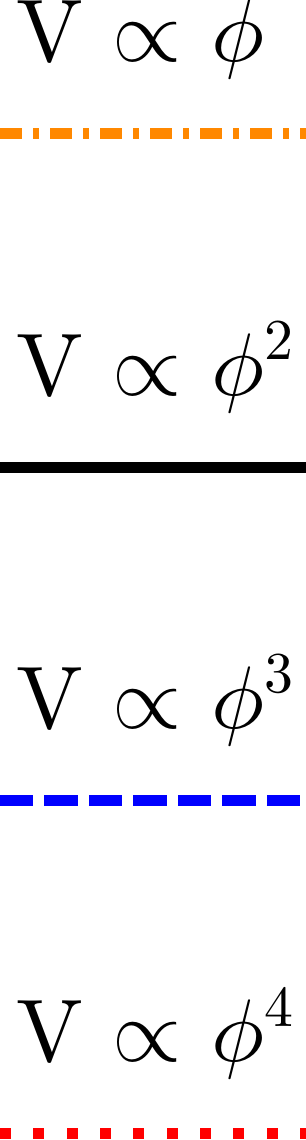}}
  \caption{ The local reconstruction of the inflationary potential. The green shaded region is the $2\sigma$ region which contains $95\%$ of the sampling
  points. In both panels, $3\times 10^4$ points are
dropped to calculate $\Delta\phi$. \\{\it Left}-- Without
using constraint on $\alpha_{\rm s}$ but instead assuming that
$\eta_2$, $\eta_3$ and $\eta_4$ have the same variance as $\eta$
(and checked that the shape of the potential does not change much
for other distributions of those variables, as long as their
variances are small).
\\ {\it Right}-- Using the constraint of $\alpha_{\rm s}$ (Eq.~(\ref{eq:alphas})) to
determine the distribution of $\eta_2$. Also assuming that
$\eta_3$ and $\eta_4$ has the same variance as $\eta$. } \label{fig:V}
\end{figure*}

Finally, with the realizations of $d_1$, $d_2$, $d_3$ and $d_4$,
we can reconstruct $V(\phi)$ locally from \eqref{eq:Vrec}. The
reconstruction is plotted in Fig. \ref{fig:V}, for $\alpha_{\rm
s}=0$ case (left panel) and large $\alpha_{\rm s}$ case (right
panel) respectively.

For the case of $\alpha_{\rm s}=0$, the reconstructed potential is
highly linear within the range $\Delta\phi\sim \Mp$. But when we
extrapolate the potential into about $10\Mp$ range, which is
suggested by large field inflation, we see higher order
derivatives does tend to bend the reconstructed potential.

In the large $\alpha_{\rm s}$ case, the reconstructed potential has significant non-linearities at
$\Delta\phi>0.2\Mp$, especially for exceptional values of $d_n$ in
the parameter space. On the other hand, we are still fine for a
local range of $\Delta\phi \simeq 0.4 \Mp$.


\begin{figure}[htbp]
  \centering
  \includegraphics[width=0.6\textwidth]{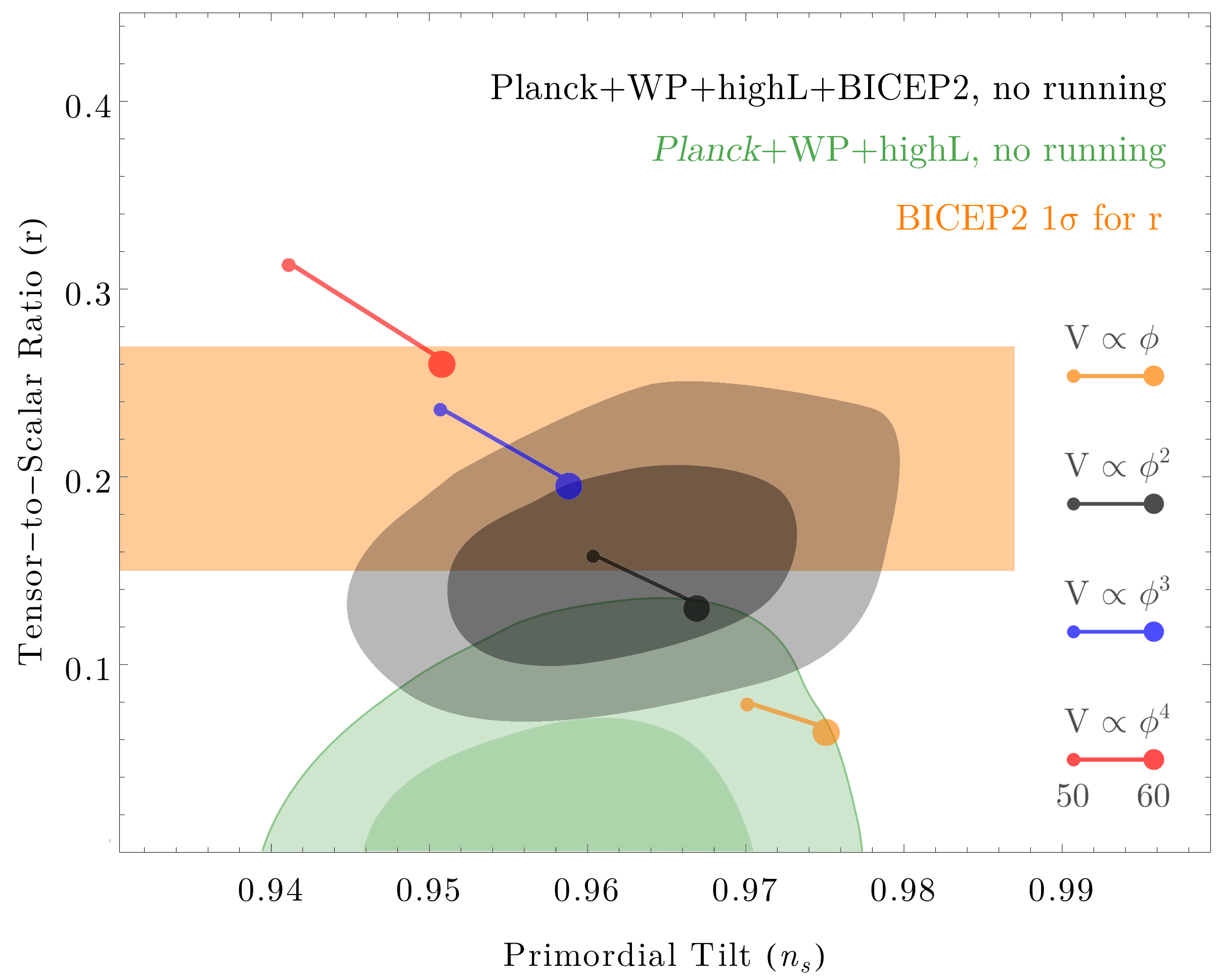}
  \caption{\label{fig:pbcombine} $n_{\rm s}$--$r$ diagram with
 predictions of $V(\phi) \sim \phi, \phi^{2},\phi^{3},\phi^{4}$
 models and the joint constraints. The \textit{Planck}+WP+highL
plots without running is extracted from
\cite{Planck22}. The BICEP2 $1\sigma$ result is taken
from \cite{BICEP2}. (For comparison, the Planck+WP+highL and Planck+WP+highL+BICEP2 results with running can be found in \cite{BICEP2}.)}
\end{figure}

\section{Comparing the large-field inflation models with data}
\label{sec:compare}

Although the local reconstruction of the inflaton potential is
safer than fitting a global potential, but nevertheless
considering the global inflationary potential by one polynomial
function is very widely used, here we compare our local reconstruction with the
global inflaton potential $V = \lambda \phi^n$.

The detailed analysis of the reheating history is beyond the scope
of the current paper. Here we use the analytical approximation
that inflation ends when $\epsilon=1$, and use slow-roll
approximation before reaching $\epsilon=1$. In this approach,
$\epsilon=1$ corresponds to $\phi_\mathrm{end} = \Mp n/\sqrt{2}$.
The relevant quantities at horizon crossing can be calculated as
\begin{align}
  \phi_* = \Mp \sqrt{2nN+\frac{n^2}{2} }~,\qquad
  \epsilon = \frac{n}{4N+n}~,\qquad
  \eta = \frac{4}{4N+n}~.
\end{align}
Those values corresponds to
\begin{align}
  P_\zeta = \frac{(4N+n) \lambda \left( \Mp \sqrt{2nN+\frac{n^2}{2} } \right)^n}{24n \pi^2 \Mp^4}~, \quad
  n_{\rm s} - 1 = - \frac{2(n+2)}{4N+n}~, \quad
  r = \frac{16n}{4N+n}~.
\end{align}
The corresponding parameters are plotted on the $n_s$-$r$ diagram in Fig.~\ref{fig:pbcombine}.

Currently, the best models which fits the BICEP2 data are the $\phi^2$ and $\phi^3$ models. From the power spectrum, $\lambda$ values for those models are (note that the power spectrum is calculated at $k=0.002 \mathrm{Mpc}^{-1}$)
\begin{align}
  \lambda = 3.1\times 10^{-11}\Mp^2 \quad (\phi^2, N=50)~,\quad
  \lambda = 1.8\times 10^{-12}\Mp \quad (\phi^3, N=50)~,
\end{align}
and
\begin{align}
  \lambda = 2.1\times 10^{-11}\Mp^2 \quad (\phi^2, N=60)~,\quad
  \lambda = 1.1\times 10^{-12}\Mp \quad (\phi^3, N=60)~.
\end{align}
Here the $\lambda$-value for the $\phi^2$ potential corresponds to
$m=7.8\times 10^{-6} \Mp$ ($N=50$) and $m=6.5\times 10^{-6} \Mp$
($N=60$). The $V\propto\phi$ and $V\propto\phi^{4}$ models can be
calculated similarly. The potential of those four models are
plotted together with the reconstructed potential in
Fig.~\ref{fig:V}.

As one can observe from the figures, the $V\propto\phi$ model
falls outside the $2\sigma$ range of reconstructed potential (in
almost whole plotted range with small $\alpha_{\rm s}$, and when
the higher derivatives not yet become dominate with large
$\alpha_{\rm s}$). While the $\lambda\phi^4$ model stays at the
boundary of $2\sigma$ at small $\Delta\phi$. This is consistent
with the $n_{\rm s}$-$r$ or $\epsilon$-$\eta$ contours in the
Fig.~\ref{fig:nsr}. On the other hand, our approach is delightful
that we now directly have the form and variance of the potential.

\section{Conclusion and discussion}
\label{sec:conclude}

We have reconstructed the inflationary potential locally around a
value $\phi_*$, which corresponds to the time when the $\ell
\simeq 50 \sim 100$ modes exits the horizon. The distribution of
the inflationary slow-roll parameters (which are defined through
the expansion) are calculated, and converted to derivatives of the
inflationary potential.

Two different assumptions have been tested against the
reconstruction -- a (theoretically) small and negligible running
of the spectral index $\alpha_{\rm s}$, and an observationally
allowed $\alpha_{\rm s}$ from current constraints. For the case of
small and negligible $\alpha_{\rm s}$, the reconstructed potential
is highly linear over $\Delta\phi \sim \Mp$ range. The effective
field theory is practically fine (although still theoretically
challenged). However, for the large $\alpha_{\rm s}$ case, higher
derivative corrections to the potential quickly dominates while
$\phi$ rolls, which implies the inflaton keeps switching
between different effective field theories, or there is a need of
a tuned inflaton field theory.

With the new observational window as shown by BICEP2 data
\cite{BICEP2}, much works are left to be done to accurately
reconstruct the amplitude and shape of the inflation potential.
Here we fit the ($n_{\rm s}$, $r$) diagram with the multi-variant
Gaussian distribution. We find that with current constraints from
{\it Planck}+WP+highL+BICEP2 data, the $V(\phi) \sim \phi^{2}$ and
$\phi^{3}$ models are consistent within $95.4\%$ CL, while $\phi$
potential is ruled out at around $99.7\%$ CL, and $\phi^{4}$ model
is consistent within $95.4\%$ CL if the number of e-folds is
around $60$. This is of-course, not a global fitting of the
inflationary prediction, but constitutes a quick examination of
the consistency between models and data.


It is also important to examine the theoretical assumption of the
shape of gravitational wave spectra. For example, if parity is
violated, which results in different amplitudes for the two tensor
modes. Another example would be non-Gaussianly distributed tensor
modes. It remains interesting to see whether the different
theoretical models can fit the new data of CMB polarization.

Theoretically, the super-Planckian range of $\phi$ motion poses
serious challenge to the field theory of inflation. It is very
important to see how to obtain theoretical naturalness for large
field inflation. Alternatively, it remains an open question that
if other sources of gravitational waves, instead of the tensor
fluctuation from the vacuum, could change the predictions.

\section*{Acknowledgments}
YZM is supported by a CITA National
Fellowship. YW is supported by a Starting Grant of the European Research Council (ERC STG grant 279617), and the Stephen Hawking Advanced Fellowship.


\end{document}